\documentclass{article}
\usepackage{spconf,amsmath,graphicx}

\usepackage{amssymb,amsfonts}
\usepackage{siunitx}
\usepackage{booktabs}  
\usepackage{placeins}
\usepackage{xurl}
\usepackage[breaklinks=true]{hyperref}

\usepackage[T1]{fontenc}
\title{Multilingual Voicemail and Answering Machine Detection via Pretrained Audio Embeddings and Gradient Boosting Machines}
\name{Kemal Altwlkany\textsuperscript{1,2}, Muamer Paric\textsuperscript{1}\thanks{This is a preprint version. The peer-reviewed article is available at:} \thanks{\url{https://doi.org/10.1007/978-3-032-37870-5_32}}}
\address{\textsuperscript{1}Infobip\\\textsuperscript{2}Faculty of Science, University of Sarajevo}

\begin{document}
%
\maketitle
\begin{abstract}
Answering machine detection (AMD) is a fundamental task in telephony that classifies whether an answered call is picked up by a human or a voicemail (machine). Despite its widespread use across cloud communication platforms, existing approaches rely on proprietary systems with undisclosed architectures, or require computationally expensive models that cannot operate in real time. We propose a lightweight AMD system that combines pretrained audio embeddings with gradient boosting machines, trained to prioritize the correct detection of human-answered calls. We evaluate seven models across two datasets: three models that we propose, a proprietary industry model, two prior academic approaches, and a large language model baseline that relies on automatic speech recognition. Furthermore, we release OpenAMD\footnote{Available at: \url{https://doi.org/10.5281/zenodo.21234058}}, the first publicly available AMD dataset comprising of more than 500 recordings in 7 languages. Our best model achieves the highest precision across both datasets while maintaining low inference latency. We assess the robustness of models to additive noise and packet loss, and perform tests of statistical significance to discover per-language and per-gender model performance. Finally, we evaluate our model on an out-of-domain task of detecting robocalls, where it achieves 94.34\% recall without any additional training, which is comparable to the industry standard.
\end{abstract}
\begin{keywords}
voicemail, answering machine, audio classification, audio embeddings, gradient boosting machine
\end{keywords}

\section{Introduction}

Answering machine detection (AMD) is a common task in telephony with the goal of classifying the first few seconds of an answered telephone call to determine who had answered the call: a human or an answering machine. The problem originates because it is not possible to distinguish answering machines from humans in telephony from signaling protocols, i.e. two common signaling protocols: Session Initiation Protocol (SIP) \cite{rfc3261} and Signaling System No.7 (SS7) \cite{itut1970ss7} do not carry information about who answered the call.

The term \textit{answering machine} is a broad term encapsulating any form of automated telephone answer and includes interactive voice responses (IVRs) and voicemails, regardless of whether they are artificially synthesized or human-recorded voicemails. The problem is framed from the caller's perspective, i.e. the caller wants to know whether the callee is a human, or an answering machine. AMD usually operates on a few seconds of the call after which the caller is provided with information about who answered the call. Based on the result, the caller can determine the further progression of the call. 

\subsection{Challenges}
The task is challenging due to several reasons. First, detection must be performed within a limited time, as phone calls are terminated in real time. If detection is not performed soon enough, there is a risk that the callee will hang up. Second, the bandwidth of telephony audio is limited; the international standard codec for telephony is G.711 \cite{itu_g711_1988}, specifically its PCMA and PCMU variants. These codecs operate at \SI{8}{\kilo\Hz} sampling rate and introduce non-linear quantization effects: 16-bit audio is compressed into 8-bit audio at a logarithmic scale. Furthermore, audio encoded with G.711 is bandlimited between \SI{300}{\Hz} and \SI{3400}{\Hz}, meaning a portion of the spectral content of speech is filtered out.
Third, the acoustic characteristics of voicemails vary depending on the callee's country, telecom operator, and language. Some voicemails are artificially synthesized messages, while some are prerecorded messages by real humans. Finally, it is often not possible to know the speaking language of the callee based on just their phone number, meaning that solutions that rely on transcribing the call must be multilingual.

\subsection{Practical Importance}

Despite its limited presence in academic literature, AMD is a core component in telephony and in modern communication platforms, having found its application across multiple tasks. We identify four principal use cases that motivate the development of an accurate AMD model.

\subsubsection{Privacy protection in anonymized calls}
Many communication platforms offer user anonymization such as number masking \cite{sinch-number-masking,telynx-number-masking,twilio-number-masking}. This is a common feature in driver, delivery, and marketplace applications. However, if a user does not answer a masked call it can reach their voicemail system and some voicemail systems might reveal the true identity to the caller, e.g.: ``You've reached 061 23...''. AMD enables platforms to detect voicemails that would reveal the true identity of the user, thereby preserving their privacy.

\subsubsection{Countermeasures for robocall fraud}
Robocalling is the practice of making large volumes of automated calls, often for fraudulent purposes. It is a well-documented problem in telecommunications security, with fraudsters stealing millions of dollars per year \cite{prasad2020s,prasad2025characterizing}. Prior research has shown that robocallers use AMD to determine when humans picked up calls in order to maximize the effectiveness of their attacks \cite{tu2016sok}. A deeper understanding of AMD can help the security community in three ways: (i) analyzing the detection techniques employed by robocallers, (ii) developing adversarial countermeasures that can deceive attacker AMD systems and thereby reduce the effectiveness of robocall attacks, and (iii) adapting AMD models for the downstream task of robocall detection itself, an approach taken in this paper as well.

\subsubsection{Public notification systems}
Government agencies, educational institutions, and emergency services rely on voice calls to deliver information. If the callee does not pick up, they can use AMD to leave messages in the callee's voicemail box such as document submission, school closers, vaccination reminders, emergency alerts, etc.

\subsubsection{Commercial campaign and call center optimization}
Voice campaigns are widely used in telemarketing, appointment reminders, and customer engagement. When a human answers the call, they can be provided with the information, but if a machine answers the call, the caller can optimize costs by not interacting with the machine, or a message can be left automatically. Cost savings can be substantial, as it is estimated that about 40-60\% of calls are answered by voicemail and in the US \cite{amazon-aws-amd}, with an estimated  2.5-4\% of the working population employed in call centers \cite{cwa-federal-call-center-factsheet,anisimovanswering}.

\subsection{Paper Contributions}

\begin{enumerate}
    \item We propose a lightweight AMD model based on pretrained audio embeddings and gradient boosting machines that prioritizes human detection rate. The model was derived through hyperparameter optimization and a model-capacity ablation where the gradient boosting component was replaced with logistic regression while keeping the remainder of the process fixed.
    \item To the best of our knowledge, we release the first AMD dataset that is publicly available. This dataset contains 233 human recordings and 300 machine recordings in 7 languages. We refer to this dataset as the OpenAMD dataset and describe it in more detail later.
    \item We compare our approach against 2 competitor models, 1 state-of-the-art proprietary model, and 1 baseline model that uses automatic speech recognition and a large language model to perform AMD. These comparisons are followed by a detailed study of the robustness and biases of AMD models, with respect to their performance across different languages and genders, as well as their robustness to noise and networking issues that can occur in practice. Finally, the performance of our model is evaluated on an out-of-domain task of detecting robocalls.
\end{enumerate}

\section{Related Work}

\begin{figure*}[!ht]
    \centering
    \includegraphics[width=0.85\linewidth]{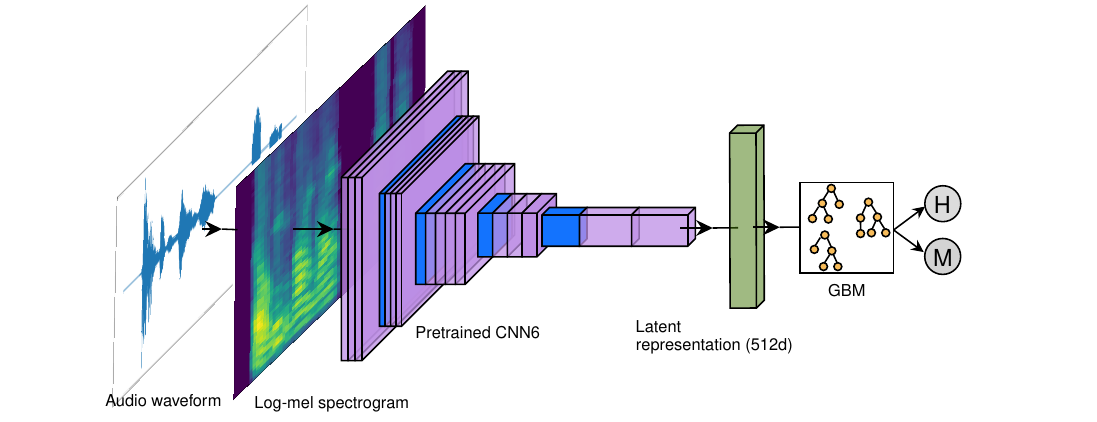}
    \caption{High-level overview of our proposed method. We compute the log-mel spectrogram from the input audio waveform and extract 512-dimensional clip-wise embeddings using PANNs CNN6 \cite{kong2020panns}. We train a GBM model on the binary classification task of AMD by using the CNN6 embeddings as inputs.}
    \label{fig:proposed-method-diagram}
\end{figure*}

Many telecommunication and cloud communication platforms provide AMD, such as Telnyx \cite{telnyx-amd}, Twilio \cite{twilio-amd}, Amazon AWS \cite{amazon-aws-amd}, Infobip \cite{infobip-amd}, ElevenLabs \cite{elevenlabs-amd}, and others. The exact implementations are unknown, with the claimed accuracies, multilingual support, inference latency, and features varying depending on the provider.

Prior to the widespread adoption of machine learning, early AMD systems used hand-crafted features and heuristics. A prominent example is Asterisk \cite{van2007asterisk,madsen2011asterisk}, an open-source private branch exchange (PBX). Asterisk's AMD \cite{Minaj2020} uses energy-based measures to detect speech and applies a temporal heuristic: short responses are labeled as human, long greetings as voicemail. This approach is simple and computationally efficient but lags behind modern models. It fails when humans use long greetings or voicemail greetings are short, and its speech detector cannot distinguish background noise from silence - any sufficiently energetic signal is treated as speech.

The approach of Gomez et al. \cite{gomez2014answering} uses automatic speech recognition (ASR) and relies on three features: (i) the time required to answer the phone, (ii) the percentage of silence in the first three seconds of the call, and (iii) the recognized sequence of words using the ASR module. These features are combined into an empirically derived function that estimates the probability that a call is answered by a human. Although this resulted in improved performance, the reliance on language-specific keywords limits generalization to multilingual settings. The system was evaluated on a small dataset of 100 phone calls with only 21 machines.

Anisimov et al. \cite{anisimovanswering} proposed the first deep learning based approach to AMD. They train a convolutional neural network (CNN) to classify calls as human or machine. The input to the CNN are mel-frequency cepstral coefficients (MFCCs) \cite{abdul2022mel,mehrish2023review} extracted from the first 2.5 seconds of the call. The authors reported an accuracy of 94\% on a dataset of 850 samples per both classes.

In \cite{altwlkany2024recurrent}, the authors addressed several of the previously reported limitations by proposing a recurrent neural network (RNN) architecture. Their approach extracted embeddings from YAMNet \cite{yamnet2020}, an audio tagging model based on the MobileNet architecture \cite{sandler2018mobilenetv2}, and fed those embeddings into a gated recurrent unit (GRU) network. The system achieved 96.7\% balanced accuracy on a dataset of approximately 4,200 calls.

Finally, \cite{bland_wav2vec_vm_finetune} is a large AMD model that was trained by fine-tuning Meta's XLS-R \cite{babu22_interspeech}, a self-supervised cross-lingual speech model that is based on wav2vec 2.0 \cite{baevski2020wav2vec}. The model contains 300 million parameters and was pretrained on more than $436\,000$  hours of speech. The AMD version was fine-tuned on English only and achieves 98\% accuracy.

\subsection{Pretrained Audio Representations}
The use of pretrained audio representations for downstream tasks has become a popular paradigm in audio and specifically speech-based machine learning applications \cite{baevski2020wav2vec,elizalde2023clap}. Models are trained on large quantities of unlabeled data and fine-tuned on small sets of labeled data.

In \cite{kong2020panns}, Kong et al. train a variety of neural networks on the task of audio tagging and achieve state-of-the-art results on several downstream audio pattern recognition tasks such as acoustic scene classification, music classification, sound event detection, etc. The authors proposed various neural architectures that rely on: conventional CNNs (architectures such as AlexNet and VGG-like), ResNets, MobileNets, one-dimensional CNNs such as DaiNet and LeeNet, and a newly proposed Wavegram-Logmel-CNN architecture. The authors jointly refer to these various networks as pretrained audio neural networks (PANNs). The CNN14 model from PANNs was used as the encoder in CLAP, a contrastive language-audio pretraining approach \cite{elizalde2023clap}. Another PANNs model, CNN6, which is based on AlexNet \cite{krizhevsky2012imagenet} achieves a mean average precision (mAP) of 0.343 on AudioSet \cite{gemmeke2017audio}, compared to their best performing Wavegram-Logmel-CNN model which achieves 0.439 mAP, but has 81 million parameters compared to the CNN6's 4.8 million.

\section{Method}

We formulate AMD as a binary classification task. Given a raw audio signal of length $T$ sampled at $f_s$, the goal is to predict a label $y \in \{\textsc{human}, \textsc{machine}\}$, with humans being the negative and machines the positive class. In practice, $T$ is bounded by a configurable detection window, typically between 1 and 5 seconds for most AMD applications \cite{telnyx-amd,twilio-amd,infobip-amd,amazon-aws-amd,elevenlabs-amd}, and $f_s$ is \SI{8}{\kilo\Hz} for narrowband telephony \cite{itu_g711_1988}.

Our approach decomposes the task into two stages: (i) a fixed feature extractor that maps the raw audio waveform into a compact embedding vector, and (ii) a lightweight classifier that operates on this embedding. This decoupled design allows the classifier to be replaced or updated independently of the feature extractor, and avoids the computational cost of end-to-end fine-tuning. A high level overview of our proposed method is provided by Figure \ref{fig:proposed-method-diagram}.

\subsection{CNN6 Audio Embeddings}

For the feature extraction stage, we use CNN6 from the PANNs \cite{kong2020panns} collection of pretrained audio models. The CNN6 model takes as input a 64-channel log-mel spectrogram; its architecture consists of four convolutional blocks with $5 \times 5$ kernels, batch normalization, ReLU activations, and average pooling, followed by two fully connected layers, totaling 4.8 million parameters.  We only use the first fully connected layer of CNN6 that produces a 512-dimensional latent representation (embedding) of the audio input. CNN6 acts as a frozen feature extractor: given an input audio segment $\mathbf{x}$, we compute the embedding $\mathbf{e} = f_{\text{CNN6}}(\mathbf{x}) \in \mathbb{R}^{512}$ without updating any network parameters.

\subsection{LightGBM Classifier}

We train a LightGBM \cite{ke2017lightgbm} classifier on the 512-dimensional CNN6 embeddings. LightGBM is a gradient boosting framework \cite{friedman2001greedy,natekin2013gradient} that constructs an ensemble of decision trees using histogram-based algorithms \cite{li2007mcrank} for efficient split finding. Compared to neural classifiers operating on the same embedding space, gradient boosting offers two advantages for the AMD use case: (i) the models produced by LightGBM are extremely lightweight, enabling real-time inference on CPU-only environments, (ii) tree-based ensembles are inherently regularized through mechanisms such as max depth, minimum leaf samples, and feature subsampling, thereby reducing the risk of overfitting on small datasets.

\subsection{Training data}

We train our GBM model on a proprietary dataset of about $3500$ manually labeled recordings, of which $800$ are human and $2700$ machines. The audio recordings are real, industry examples of AMD and are already transcoded via G.711 \cite{itu_g711_1988}, with the presence of background and static noise. Although most of the recordings are in English, the dataset does contain recordings in other languages.

To the best of our knowledge, we were unable to find any publicly available datasets for AMD. This comes to no surprise given the high sensitivity of the topic from a user privacy perspective, making it difficult to record and publish data. The lack of publicly available AMD datasets makes it difficult to reproduce and verify the models.

\subsubsection{OpenAMD Dataset}

We introduce the OpenAMD dataset, a multilingual AMD dataset that contains realistic recordings of answered phone calls, both by humans and by machines, comprising 533 recordings in total.

The human samples in the dataset were obtained by recording phone calls from users that participated in the study, as we found the recordings to be more natural when participants would answer the phone instead of speak into a recording device in a studio environment. In total we have 8 different speakers in 6 languages.

We used Azure Cognitive Services Neural TTS \cite{azure-speech-studio} to synthesize machine samples in 7 languages (English, Chinese (Mandarin), Arabic, German, etc.), containing a total of 44 different voices, balanced in terms of gender.

Since the voicemails are neurally synthesized, we simulate realistic telephony conditions by introducing augmentations such as adding ringing tones, silent gaps and pauses, and simulating pick-up clicks of calls being connected.  Furthermore, we downsample all audio to \SI{8}{\kilo\Hz} and transcode it using G.711 \cite{itu_g711_1988}, randomly altering between PCMA and PCMU. This process introduces the non-linear quantization artifacts, bandwidth limitation, and dynamic range compression characteristic of telephony-grade audio, making the synthetic samples more acoustically similar to real recordings in the training set.

We make the OpenAMD dataset publicly available and accompany the audio recordings with useful metadata: speaker ID (or neural voice for machines), speaker gender, recording text (for both humans and machines), and recording language.

\begin{table*}[!ht]
    \centering
    \caption{Performance of AMD models on proprietary dataset.}
    \resizebox{0.75\linewidth}{!}{%
\centering
  \begin{tabular}{lccccc}
  \toprule
  Model & Accuracy & Precision & Recall & Specificity & F1 \\
  \midrule
  RNN-YAMNet \cite{altwlkany2024recurrent}         & $83.35 \pm 2.25$ & $87.72 \pm 3.04$ & $74.40 \pm 4.04$ & $91.04 \pm 2.43$ & $80.49 \pm 2.56$ \\
  AMD-Infobip \cite{infobip-amd}          & $88.09 \pm 1.86$ & $92.09 \pm 2.59$ & $81.20 \pm 3.55$ & $94.01 \pm 1.96$ & $86.29 \pm 2.44$ \\
  CNN6-LogReg (our)         & $85.61 \pm 2.38$ & $91.55 \pm 2.65$ & $75.87 \pm 4.43$ & $93.99 \pm 1.89$ & $82.96 \pm 2.73$ \\
  CNN6-GBM-balanced (our)    & $85.91 \pm 2.46$ & $92.33 \pm 2.95$ & $75.83 \pm 3.83$ & $94.59 \pm 2.27$ & $83.25 \pm 3.03$ \\
  CNN6-GBM (our)           & $85.20 \pm 2.10$ & $\mathbf{94.70} \pm 2.37$ & $72.02 \pm 4.08$ & $\mathbf{96.54} \pm 1.49$ & $81.80 \pm 2.81$ \\
  wav2vec-vm-fine \cite{bland_wav2vec_vm_finetune}      & $84.72 \pm 2.21$ & $87.77 \pm 2.99$ & $77.79 \pm 3.87$ & $90.68 \pm 2.13$ & $82.46 \pm 2.83$ \\
  whisper-GPT5.4 (baseline) & $\mathbf{91.09} \pm 1.59$ & $86.58 \pm 3.40$ & $\mathbf{95.53} \pm 1.86$ & $87.28 \pm 2.77$ & $\mathbf{90.83} \pm 1.80$ \\
  \bottomrule
  \end{tabular}
    }
    \label{tab:performance-on-proprietary-dataset}
\end{table*}

\begin{figure*}[ht]
    \centering
    \includegraphics[width=0.95\linewidth]{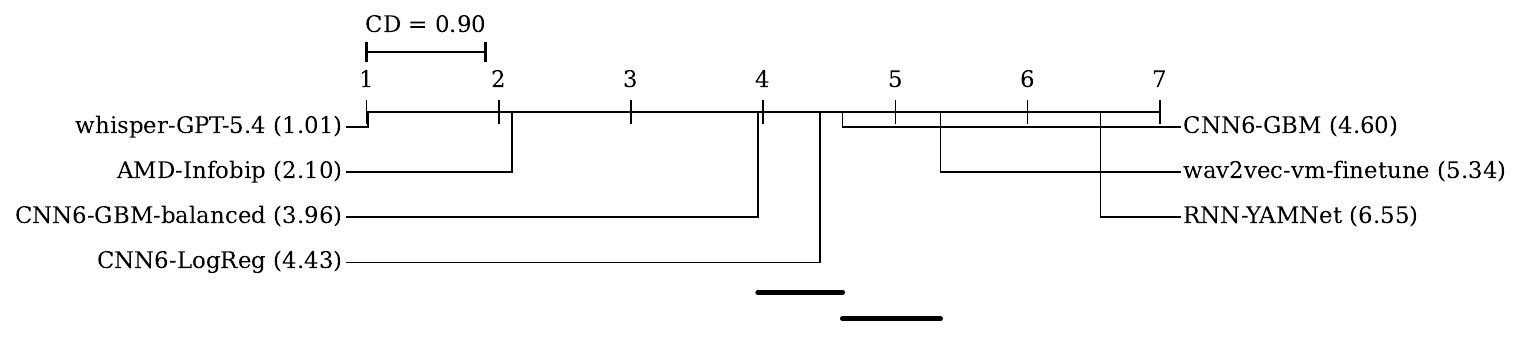}
    \caption{Critical difference diagram showing the average ranks of the models across \textit{accuracy}.}
    \label{fig:critical-difference-diagram-accuracy}
\end{figure*}

\subsection{Experimental setup}

All results that we present are aggregates over 5 repeated experiment runs and are computed at a 95\% confidence interval using the bootstrap approach \cite{Hastie09-TEO}. Experiments were performed on an NVIDIA DGX B200 platform.

\subsubsection{Baseline models}
We establish a high-quality baseline AMD model that transcribes calls using ASR and uses an LLM to classify them based on the transcript as human or machine. The transcription model needs to be a large multilingual model, and the LLM needs to be sufficiently powerful to understand the task at hand. We opted for Whisper \cite{radford2023robust} (large-v3-turbo model) and for ChatGPT5.4. Whisper has close to 800 million parameters, while the exact number of parameters of ChatGPT-5.4 is undisclosed, but widely estimated to surpass 2 trillion parameters. This approach does therefore not scale well and would not be of practical use, as AMD must be a resource efficient model that can run in real time on a CPU. However, it can be beneficial to use this approach as a baseline given its large predictive capability.

Regarding existing approaches, we compare to \textit{RNN-YAMNet} \cite{altwlkany2024recurrent} and \textit{wav2vec-vm-finetune} \cite{bland_wav2vec_vm_finetune}, and a proprietary state-of-the-art model \textit{AMD-Infobip} \cite{infobip-amd}.

\subsubsection{GBM models}
We hypothesize that, given the AMD use cases described in the introductory section, it is of paramount importance for the model to classify humans correctly. In particular, the model should prioritize precision and specificity, since falsely classifying humans as machines may lead to worse outcomes than falsely classifying machines as humans. Therefore, we provide three model versions:
\begin{itemize}
    \item \textit{CNN6-GBM}, our primary model, prioritizes precision and specificity, thus minimizing the chance of wrongly classifying humans
    \item \textit{CNN6-GBM-balanced}, our balanced model that provides a good trade-off between recall and specificity
    \item \textit{CNN6-LogReg}, a model trained using logistic regression instead of GBM, with the goal of investigating the contribution of the non-linear gradient boosting component and whether CNN6 produces linearly separable embeddings
\end{itemize}

\section{Results}

\subsection{Proprietary dataset}
Table \ref{tab:performance-on-proprietary-dataset} provides the benchmarks of all 7 models on a proprietary dataset (different from the training set). The whisper-GPT-5.4 baseline provides the best balanced performance such as accuracy and F1-score, but the CNN6-GBM model has the best specificity and precision.

We compared the \textit{accuracy} of the 7 classifiers using the Friedman test which showed a statistically significant difference among classifiers ($\chi^{2}(6) = 448.78,\,\, p \ll 1e\text{-}10$), with Kendall's $W = .748$, indicating a large degree of consistency in classifier rankings. Post-hoc Nemenyi tests with Holm-Bonferroni correction were used to compare average ranks across classifiers and showed that models CNN6-GBM, CNN6-GBM-balanced, CNN6-LogReg, and wav2vec-vm-finetune do not differ significantly in terms of accuracy, whereas whisper-GPT-5.4 and AMD-Infobip are significantly better than other models, while RNN-YAMNet is the worst ranked model. This is illustrated by the critical difference diagram in Figure \ref{fig:critical-difference-diagram-accuracy}.

\begin{figure*}[ht]
    \centering
    \includegraphics[width=0.95\linewidth]{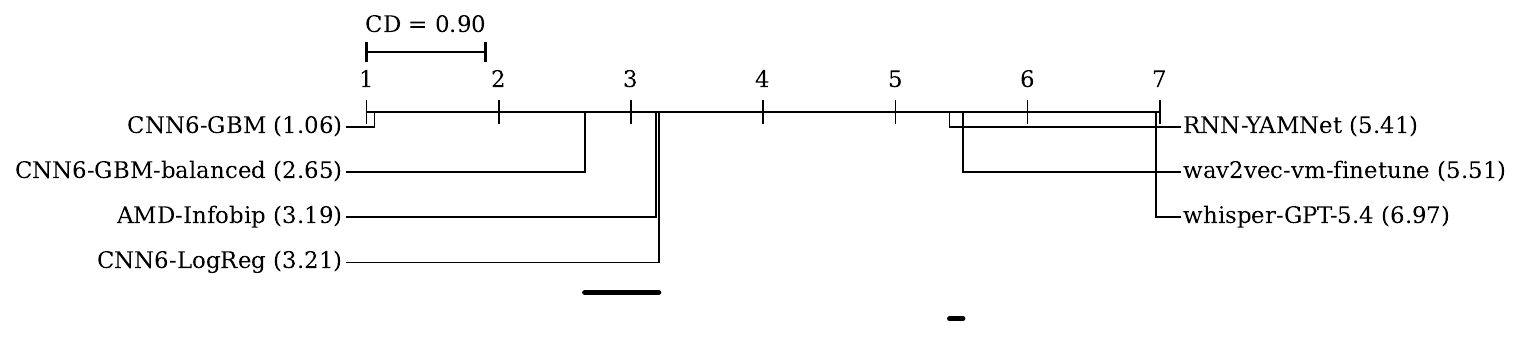}
    \caption{Critical difference diagram showing the average ranks of the models across \textit{specificity}.}
    \label{fig:critical-difference-diagram-specificity}
\end{figure*}

We repeated the procedure but analyzed \textit{specificity} among the 7 classifiers and obtained: ($\chi^{2}(6) = 523.04,\,\, p \ll 1e\text{-}10$), with Kendall's $W = .872$, indicating an even larger degree of consistency in classifier rankings among specificity compared to accuracy. Post-hoc Nemenyi tests with Holm-Bonferroni correction showed that our CNN6-GBM differs significantly from other models, while there is no statistically significant difference between the performance of AMD-Infobip and CNN6-LogReg, and between RNN-YAMNet and wav2vec-vm-finetune. This is illustrated in Figure \ref{fig:critical-difference-diagram-specificity}.

\begin{table*}
    \centering
    \caption{Performance of AMD models on OpenAMD dataset.}
    \resizebox{0.85\linewidth}{!}{%
\centering
\begin{tabular}{llllll}
\toprule
Model & Accuracy & Precision & Recall & Specificity & F1 \\
\midrule
RNN-YAMNet \cite{altwlkany2024recurrent} & \textbf{94.56} $\pm$ 2.06 & 91.69 $\pm$ 3.38 & \textbf{99.33} $\pm$ 1.06 & 88.41 $\pm$ 4.36 & \textbf{95.36} $\pm$ 1.88 \\
AMD-Infobip \cite{infobip-amd} & 91.74 $\pm$ 2.44 & 88.55 $\pm$ 3.64 & 98.00 $\pm$ 1.82 & 83.69 $\pm$ 4.98 & 93.04 $\pm$ 2.19 \\
CNN6-LogReg (our)  & 88.74 $\pm$ 2.81 & 87.97 $\pm$ 3.73 & 92.67 $\pm$ 3.14 & 83.69 $\pm$ 4.72 & 90.26 $\pm$ 2.70 \\
CNN6-GBM-balanced (our) & 91.93 $\pm$ 2.44 & 92.41 $\pm$ 3.12 & 93.33 $\pm$ 2.94 & 90.13 $\pm$ 4.04 & 92.87 $\pm$ 2.26 \\
CNN6-GBM (our) & 92.50 $\pm$ 2.44 & \textbf{95.14} $\pm$ 2.64 & 91.33 $\pm$ 3.51 & 93.99 $\pm$ 3.12 & 93.20 $\pm$ 2.38 \\
wav2vec-vm-finetune \cite{bland_wav2vec_vm_finetune} & 84.43 $\pm$ 3.00 & 95.02 $\pm$ 2.97 & 76.33 $\pm$ 4.72 & \textbf{94.85} $\pm$ 2.99 & 84.66 $\pm$ 3.23 \\
whisper-GPT-5.4 (baseline) & 91.56 $\pm$ 2.44 & 87.61 $\pm$ 3.52 & 99.00 $\pm$ 1.30 & 81.97 $\pm$ 5.05 & 92.96 $\pm$ 2.08 \\
\bottomrule
\end{tabular}
    }
    \label{tab:performance-on-open-source-dataset}
\end{table*}

\begin{figure*}[ht]
    \centering
    \includegraphics[width=0.95\linewidth]{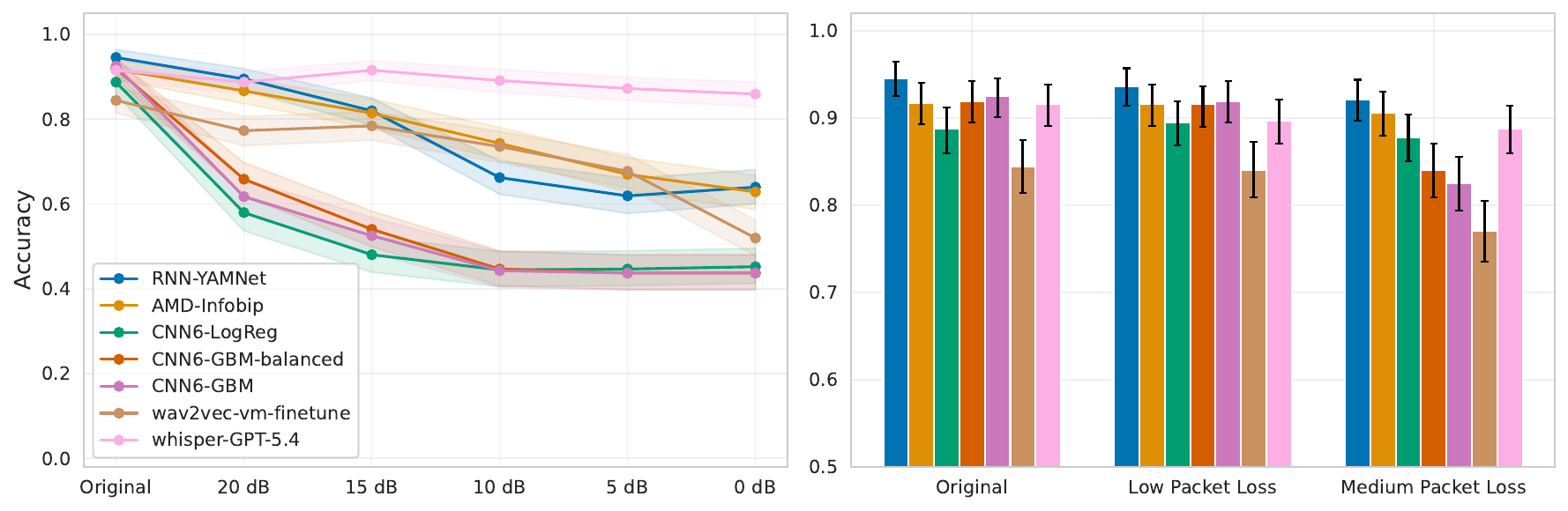}
    \caption{Accuracy of AMD models in noisy and poor network quality conditions. Left: audio is distorted by noise (x-axis indicates SNR). Right: audio is corrupted by packet loss. Shaded areas and black bars indicate the 95\% CI.}
    \label{fig:combined-robustness}
\end{figure*}

\subsection{OpenAMD dataset}

We now benchmark the models on the OpenAMD dataset and provide the results in Table \ref{tab:performance-on-open-source-dataset}. The ranks of the models differ compared to the proprietary dataset, with the exception of CNN6-GBM being the best performing model in terms of precision on both datasets. On the OpenAMD dataset, RNN-YAMNet obtains the best accuracy, recall, and F1 scores, with wav2vec-vm-finetune now achieving the best specificity.

\subsubsection{Robustness to noise and coding artifacts}

In Figure \ref{fig:combined-robustness} we show the performance of the models during non-ideal conditions: when the audio has been corrupted with noise or when packet loss is present in the network (simulated using the Gilbert-Elliot model \cite{pieper2023relationships}). The results indicate that the whisper-GPT-5.4 model is most robust to noise distortions, whereas the GBM-based models are the worst performing models for noisy audio. The GBM-based models maintain good performance in poor network quality conditions, with RNN-YAMNet and AMD-Infobip being the most robust models for packet loss.

\subsubsection{Per-language performance}
We analyzed whether the performance of the models depends on the language of the callee/voicemail. A chi-squared test of independence was performed on each model's accuracy across different languages, with Holm-Bonferroni correction for multiple comparisons. Six of the models showed no statistically significant language dependence, i.e. the accuracies of the models are independent of the language. For wav2vec-vm-finetune we obtained ($\chi^{2}(4) = 20.28, \,\, p = 0.003$) with Cramer's $V=0.218$ indicating a statistically significant but small association, i.e. the model's performance depends on the language, but not strongly. We hypothesize that this might be due to the model being fine-tuned on English recordings only, thereby degrading the previously learned knowledge of the XLS-R model \cite{babu22_interspeech}.

\subsubsection{Per-gender performance}

\begin{figure*}[ht]
    \centering
    \includegraphics[width=0.95\linewidth]{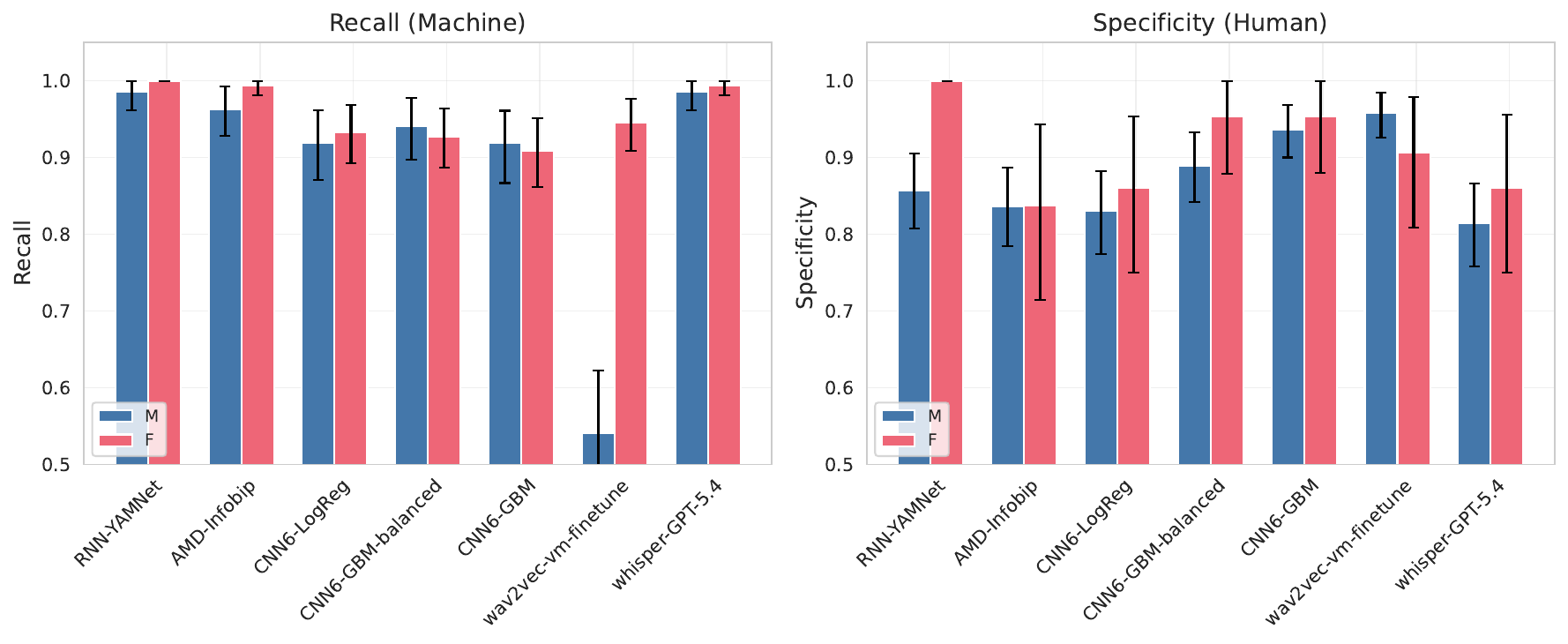}
    \caption{Machine recall and human specificity, per gender and per model. Black bars indicate the 95\% CI.}
    \label{fig:gender-analysis}
\end{figure*}

We ran a separate Fisher's exact test for machine recall and for human specificity to see if male vs. female recordings of humans/machines get correctly detected at the same rate. The recall and specificity per each model are provided by Figure \ref{fig:gender-analysis}. For machine recall, only wav2vec-vm-fintune exhibited a significant gender disparity (odds ratio $OR = .068,\,\, p < .001$) indicating that male voicemails are $\approx15$ times as likely to be misclassified as human. For human specificity, no statistically significant results were obtained, although RNN-YAMnet provides a relatively large gap between the performance in male vs. female specificity.

\subsection{Inference latency}

In Table \ref{tab:latency} we show the inference latency of the CNN6-based models and wav2vec-vm-finetune. Note that we did not have access to the inference latency information of the RNN-YAMNet, AMD-Infobip, and whisper-GPT-5.4 models, but it is clear that the latter cannot operate in real-time given its reliance on Whisper \cite{radford2023robust} and ChatGPT-5.4. The average latency of CNN6-based models is two orders of magnitude less than that of the wav2vec-vm-finetune model, with the balanced GBM model having a more variable inference latency compared to other CNN6-based models.

\begin{table}[!ht]
    \centering
    \caption{Inference latency of AMD models (in milliseconds).}
    \resizebox{0.8\linewidth}{!}{%
\centering
  \begin{tabular}{lrrr}
    \toprule
    Model     & Mean          & Median      & Std  \\
    \midrule
    CNN6-GBM               &1.23 & 1.21   & 0.092\\
    CNN6-GBM-balanced      &1.36 & 1.33   & 0.358\\
    CNN6-LogReg            &1.19 & 1.18   & 0.062\\
    wav2vec-vm-finetune    &131.99 & 132.85   & 14.179\\
    \bottomrule
  \end{tabular}
    }
    \label{tab:latency}
\end{table}

\subsection{Ablation study}

We performed a grid search on the parameters of CNN6-GBM, summarized in Table \ref{tab:grid-search}. From the results it appears that our model has found ``sweet spots''; optimal parameter configurations are located ``in the middle'', with no parameter taking extreme values. Increasing the number of estimators beyond 200 yields no positive gains, nor does increasing the number of leaves beyond 61.
Additionally, we argue that the non-linearity of gradient boosting results in a small but noticeable performance improvement as opposed to logistic regression, as can be inferred by comparing the performances of the CNN6-GBM and CNN6-LogReg models. CNN6-LogReg performs worse than CNN6-GBM in terms of accuracy, precision, and F1-score on the OpenAMD dataset, with the difference in specificity greater than 10\% on average. Given that the AMD-LogReg inference latency is on average \SI{0.04}{\milli\second} faster, we argue that the performance boost provided by GBM is worth the negligible increase in computational complexity.

\begin{table}
    \centering
    \caption{CNN6-GBM hyperparameter search space and selected values.}
    \resizebox{1.0\linewidth}{!}{%
\centering
  \begin{tabular}{lll}
    \toprule
    Parameter     & Search space          & Selected \\
    \midrule
    Learning rate          & \{0.03, 0.1, 0.3\}      & 0.1 \\
    Number of estimators   & \{50, 100, 200, 300, 400, 500\}        & 200 \\
    Num leaves             & \{17, 31, 41, 51, 61, 127\}           & 61 \\
    Min child samples      & \{5, 10, 20, 30, 50, 100, 200\}               & 50 \\
    \bottomrule
  \end{tabular}
    }
    \label{tab:grid-search}
\end{table}

\subsection{Transfer to Out-of-Domain Task}
We use the CNN6-GBM model as-is and apply it to an out-of-domain task of detecting robocalls \cite{tu2016sok,prasad2020s,prasad2025characterizing} without any additional training or fine-tuning. We use the dataset from the US Federal Trade Commission (FTC) \cite{prasad2023robocall} that contains 1432 robocalls (all positive samples) and manage to detect 1351 robocalls, which is a recall rate of 94.34\%, close to the industry standard of 93-97\% \cite{elizalde2021detection,pandit2023combating}. We hope that the performance of CNN6-GBM could be further improved by additional training so that it could be used to detect fraudulent robocalls.

\subsection{Discussion}

We proposed three CNN6-based models, with significantly lower inference time than competitor approaches. Our best performing model CNN6-GBM, tuned to prioritize the rate of human detection provided the best precision on both datasets and is statistically the best performing model in terms of specificity on the first dataset. We argue that the high precision rate is due to our goal of minimizing false positives; we want to ensure that the model is very confident before it classifies a call as voicemail, thereby reducing the chances of falsely categorizing humans as machines. Removing the non-linear gradient boosting component impacts performance as indicated by the CNN6-LogReg model, while yielding minimal improvements in inference latency. The hyperparameter optimization study showed that increasing the number of estimators or leaves improves the model's predictive capabilities up to a certain point, after which no improvement gains were seen.
There is room for improvement regarding noise robustness, as the CNN6-based models perform worse than competitors in this area, although they maintain decent performance during poor network conditions. The CNN6-based models showed no bias in terms of language or gender, implying that the models learned to generalize well and can be used in multilingual settings, regardless of the language or gender of the speaker.
\section{Conclusion}

In this paper we showed a new approach to AMD by using pretrained audio neural networks and combining them with gradient boosting machines. Our model reaches comparable accuracy to the industry standard and current state-of-the-art, it is the best performing model in terms of precision while maintaining high specificity, thereby maximizing the correct detection rate of humans. We created OpenAMD, a multilingual and diverse dataset and make it publicly available. Our analysis shows that our models generalize well and are not biased to any language or gender. The robustness to packet loss is good, although there is room for improvement regarding robustness to additive noise. Finally, our model showed to perform well on the out-of-domain task of robocall detection without any additional training or fine-tuning.

\bibliographystyle{IEEEbib}
\bibliography{references}

@techreport{itut1970ss7,
  author      = {{International Telecommunication Union}},
  title       = {{Introduction to CCITT Signalling System No. 7}},
  type        = {ITU-T Recommendation},
  number      = {Q.700},
  institution = {International Telecommunication Union},
  address     = {Geneva, Switzerland},
  month       = nov,
  year        = {1988},
  url         = {https://www.itu.int/rec/T-REC-Q.700}
}

@techreport{itu_g711_1988,
  author      = {{International Telecommunication Union}},
  title       = {Pulse code modulation ({PCM}) of voice frequencies},
  type        = {ITU-T Recommendation},
  number      = {G.711},
  institution = {International Telecommunication Union},
  address     = {Geneva, Switzerland},
  month       = nov,
  year        = {1988},
  url         = {https://www.itu.int/rec/T-REC-G.711-198811-I/en}
}

@misc{rfc3261,
    series =    {Request for Comments},
    number =    3261,
    howpublished =  {RFC 3261},
    publisher = {RFC Editor},
    doi =       {10.17487/RFC3261},
    url =       {https://www.rfc-editor.org/info/rfc3261},
    author =    {Eve Schooler and Jonathan Rosenberg and Henning Schulzrinne and Alan Johnston and Gonzalo Camarillo and Jon Peterson and Robert Sparks and Mark J. Handley},
    title =     {{SIP: Session Initiation Protocol}},
    pagetotal = 269,
    year =      2002,
    month =     jul,
}

@article{abdul2022mel,
  title={Mel frequency cepstral coefficient and its applications: A review},
  author={Abdul, Zrar Kh and Al-Talabani, Abdulbasit K},
  journal={Ieee Access},
  volume={10},
  pages={122136--122158},
  year={2022},
  publisher={IEEE}
}

@article{mehrish2023review,
  title={A review of deep learning techniques for speech processing},
  author={Mehrish, Ambuj and Majumder, Navonil and Bharadwaj, Rishabh and Mihalcea, Rada and Poria, Soujanya},
  journal={Information Fusion},
  volume={99},
  pages={101869},
  year={2023},
  publisher={Elsevier}
}

@mastersthesis{Minaj2020,
  author  = {Minaj, Enxhi},
  title   = {Optimization of Answering Machine Detection Application in Asterisk},
  school  = {Lappeenranta-Lahti University of Technology (LUT University)},
  year    = {2020},
  address = {Lappeenranta, Finland},
  url     = {https://lutpub.lut.fi/handle/10024/161698}
}

@misc{bland_wav2vec_vm_finetune,
  author       = {Downie, Jake},
  title        = {{wav2vec-vm-finetune}},
  year         = {2025},
  howpublished = {\url{https://huggingface.co/jakeBland/wav2vec-vm-finetune}},
  note = {[Online] Accessed: 2026-05-11}
}

@inproceedings{altwlkany2024recurrent,
  title={A Recurrent Neural Network Approach to the Answering Machine Detection Problem},
  author={Altwlkany, Kemal and Delali{\'c}, Sead and Selmanovi{\'c}, Elmedin and Alihod{\v{z}}i{\'c}, Adis and Lovri{\'c}, Ivica},
  booktitle={2024 47th MIPRO ICT and Electronics Convention (MIPRO)},
  pages={85--90},
  year={2024},
  organization={IEEE}
}

@inproceedings{tu2016sok,
  title={Sok: Everyone hates robocalls: A survey of techniques against telephone spam},
  author={Tu, Huahong and Doup{\'e}, Adam and Zhao, Ziming and Ahn, Gail-Joon},
  booktitle={2016 IEEE Symposium on Security and Privacy (SP)},
  pages={320--338},
  year={2016},
  organization={IEEE}
}

@inproceedings{gomez2014answering,
  title={Answering machine detector for call centers},
  author={G{\'o}mez, Jon A and Montes-Novella, Luis M and Garc{\i}a, Jos{\'e}},
  booktitle={Proc. iberSPEECH 2014},
  pages={198--207},
  year={2014}
}

@article{anisimovanswering,
  title={Answering machines detection in contact centers using convolutional neural networks},
  author={Anisimov, Nikolay and Olkhovsky, Mikhail and Kishinsky, Konstantin},
  journal={FizTech-USA},
  year={2023}
}

@article{kong2020panns,
  title={Panns: Large-scale pretrained audio neural networks for audio pattern recognition},
  author={Kong, Qiuqiang and Cao, Yin and Iqbal, Turab and Wang, Yuxuan and Wang, Wenwu and Plumbley, Mark D},
  journal={IEEE/ACM Transactions on Audio, Speech, and Language Processing},
  volume={28},
  pages={2880--2894},
  year={2020},
  publisher={IEEE}
}

@article{elizalde2021detection,
    author = {Elizalde, Benjamin and Emmanouilidou, Dimitra},
    title = {Detection of Robocall and Spam Calls using Acoustic Features of Incoming Voicemails},
    journal = {Proceedings of Meetings on Acoustics},
    volume = {45},
    number = {1},
    pages = {060004},
    year = {2022},
    month = {02},
    issn = {1939-800X},
    doi = {10.1121/2.0001533},
}

@article{prasad2023robocall,
  title={Robocall Audio from the FTC’s Project Point of No Entry},
  author={Prasad, Sathvik and Reaves, Bradley},
  journal={North Carolina State University, Tech. Rep. TR-2023-1},
  year={2023}
}

@inproceedings{prasad2020s,
  title={Who's calling? characterizing robocalls through audio and metadata analysis},
  author={Prasad, Sathvik and Bouma-Sims, Elijah and Mylappan, Athishay Kiran and Reaves, Bradley},
  booktitle={29th USENIX Security Symposium (USENIX Security 20)},
  pages={397--414},
  year={2020}
}

@inproceedings{radford2023robust,
  title={Robust speech recognition via large-scale weak supervision},
  author={Radford, Alec and Kim, Jong Wook and Xu, Tao and Brockman, Greg and McLeavey, Christine and Sutskever, Ilya},
  booktitle={International conference on machine learning},
  pages={28492--28518},
  year={2023},
  organization={PMLR}
}

@misc{yamnet2020,
  author = {{Google}},
  title = {{YAMNet: A deep net for audio classification}},
  howpublished = {\url{https://github.com/tensorflow/models/tree/master/research/audioset/yamnet}},
  note = {[Online] Accessed: 2026-05-11},
  year = {2020}
}

@book{madsen2011asterisk,
  title={Asterisk: The definitive guide},
  author={Madsen, Leif and Van Meggelen, Jim and Bryant, Russell},
  year={2011},
  publisher={" O'Reilly Media, Inc."}
}

@book{van2007asterisk,
  title={Asterisk: The future of telephony},
  author={Van Meggelen, Jim and Madsen, Leif and Smith, Jared},
  year={2007},
  publisher={" O'Reilly Media, Inc."}
}

@inproceedings{elizalde2023clap,
  title={Clap learning audio concepts from natural language supervision},
  author={Elizalde, Benjamin and Deshmukh, Soham and Al Ismail, Mahmoud and Wang, Huaming},
  booktitle={ICASSP 2023-2023 IEEE International Conference on Acoustics, Speech and Signal Processing (ICASSP)},
  pages={1--5},
  year={2023},
  organization={IEEE}
}

@inproceedings{gemmeke2017audio,
  title={Audio set: An ontology and human-labeled dataset for audio events},
  author={Gemmeke, Jort F and Ellis, Daniel PW and Freedman, Dylan and Jansen, Aren and Lawrence, Wade and Moore, R Channing and Plakal, Manoj and Ritter, Marvin},
  booktitle={2017 IEEE international conference on acoustics, speech and signal processing (ICASSP)},
  pages={776--780},
  year={2017},
  organization={IEEE}
}

@article{krizhevsky2012imagenet,
  title={Imagenet classification with deep convolutional neural networks},
  author={Krizhevsky, Alex and Sutskever, Ilya and Hinton, Geoffrey E},
  journal={Advances in neural information processing systems},
  volume={25},
  year={2012}
}

@inproceedings{sandler2018mobilenetv2,
  title={Mobilenetv2: Inverted residuals and linear bottlenecks},
  author={Sandler, Mark and Howard, Andrew and Zhu, Menglong and Zhmoginov, Andrey and Chen, Liang-Chieh},
  booktitle={Proceedings of the IEEE conference on computer vision and pattern recognition},
  pages={4510--4520},
  year={2018}
}

@article{baevski2020wav2vec,
  title={wav2vec 2.0: A framework for self-supervised learning of speech representations},
  author={Baevski, Alexei and Zhou, Yuhao and Mohamed, Abdelrahman and Auli, Michael},
  journal={Advances in neural information processing systems},
  volume={33},
  pages={12449--12460},
  year={2020}
}

@article{friedman2001greedy,
  title={Greedy function approximation: a gradient boosting machine},
  author={Friedman, Jerome H},
  journal={Annals of statistics},
  pages={1189--1232},
  year={2001},
  publisher={JSTOR}
}

@article{natekin2013gradient,
  title={Gradient boosting machines, a tutorial},
  author={Natekin, Alexey and Knoll, Alois},
  journal={Frontiers in neurorobotics},
  volume={7},
  pages={63623},
  year={2013},
  publisher={Frontiers}
}

@inproceedings{ke2017lightgbm,
  title={{LightGBM}: A Highly Efficient Gradient Boosting Decision Tree},
  author={Ke, Guolin and Meng, Qi and Finley, Thomas and Wang, Taifeng and Chen, Wei and Ma, Weidong and Ye, Qiwei and Liu, Tie-Yan},
  booktitle={Advances in Neural Information Processing Systems (NeurIPS)},
  volume={30},
  pages={3146--3154},
  year={2017}
}

@inproceedings{babu22_interspeech,
  title     = {{XLS-R: Self-supervised Cross-lingual Speech Representation Learning at Scale}},
  author    = {Arun Babu and Changhan Wang and Andros Tjandra and Kushal Lakhotia and Qiantong Xu and Naman Goyal and Kritika Singh and Patrick {von Platen} and Yatharth Saraf and Juan Pino and Alexei Baevski and Alexis Conneau and Michael Auli},
  year      = {2022},
  booktitle = {{Interspeech 2022}},
  pages     = {2278--2282},
  doi       = {10.21437/Interspeech.2022-143},
  issn      = {2958-1796},
}

@book{Hastie09-TEO,
    address      = {New York},
    author       = {{Trevor Hastie and Robert Tibshirani and Jerome Friedman}},
    publisher    = {{Springer}},
    title        = {{The Elements of Statistical Learning -- Data Mining, Inference, and Prediction}},
    year         = 2009,
}

@techreport{pieper2023relationships,
  title={Relationships between Gilbert-Elliot Burst Error Model Parameters and Error Statistics},
  author={Pieper, Jaden},
  year={2023},
  institution={Institute for Telecommunication Sciences}
}

@inproceedings{prasad2025characterizing,
  title={Characterizing robocalls with multiple vantage points},
  author={Prasad, Sathvik and Nahapetyan, Aleksandr and Reaves, Bradley},
  booktitle={2025 IEEE Symposium on Security and Privacy (SP)},
  pages={19--36},
  year={2025},
  organization={IEEE}
}

@article{li2007mcrank,
  title={Mcrank: Learning to rank using multiple classification and gradient boosting},
  author={Li, Ping and Wu, Qiang and Burges, Christopher},
  journal={Advances in neural information processing systems},
  volume={20},
  year={2007}
}

@misc{infobip-amd,
  author       = {Infobip},
  title        = {{Automated machine detection}},
  year         = {2026},
  howpublished = {\url{https://www.infobip.com/docs/voice-and-video/calls#automated-machine-detection-understanding-calls-api}},
  note = {[Online] Accessed: 2026-05-11}
}

@misc{telnyx-amd,
  author       = {Telnyx},
  title        = {{Answering Machine Detection}},
  year         = {2026},
  howpublished = {\url{https://developers.telnyx.com/docs/voice/programmable-voice/answering-machine-detection}},
  note = {[Online] Accessed: 2026-05-11}
}

@misc{amazon-aws-amd,
  author       = {{Amazon AWS}},
  title        = {{Best practices for answering machine detection}},
  year         = {2026},
  howpublished = {\url{https://docs.aws.amazon.com/connect/latest/adminguide/outbound-campaign-best-practices.html}},
  note = {[Online] Accessed: 2026-05-11}
}

@misc{twilio-amd,
  author       = {{Twilio}},
  title        = {{Answering Machine Detection}},
  year         = {2026},
  howpublished = {\url{https://www.twilio.com/docs/voice/answering-machine-detection}},
  note = {[Online] Accessed: 2026-05-11}
}

@misc{elevenlabs-amd,
  author       = {{ElevenLabs}},
  title        = {{Voicemail detection}},
  year         = {2026},
  howpublished = {\url{https://elevenlabs.io/docs/eleven-agents/customization/tools/system-tools/voicemail-detection}},
  note = {[Online] Accessed: 2026-05-11}
}

@misc{telynx-number-masking,
  author       = {{Telnyx}},
  title        = {{Phone number masking}},
  year         = {2026},
  howpublished = {\url{https://telnyx.com/use-cases/phone-number-masking}},
  note = {[Online] Accessed: 2026-05-11}
}

@misc{sinch-number-masking,
  author       = {{Sinch}},
  title        = {{Protect privacy with number masking}},
  year         = {2026},
  howpublished = {\url{https://sinch.com/voice/programmable-voice/number-masking/}},
  note = {[Online] Accessed: 2026-05-11}
}

@misc{twilio-number-masking,
  author       = {{Twilio}},
  title        = {{What are Masked Phone Numbers?}},
  year         = {2026},
  howpublished = {\url{https://www.twilio.com/docs/glossary/what-are-masked-phone-numbers}},
  note = {[Online] Accessed: 2026-05-11}
}

@misc{azure-speech-studio,
  author       = {{Microsoft}},
  title        = {{Azure Speech Studio}},
  year         = {2026},
  howpublished = {\url{https://speech.microsoft.com/}},
  note = {[Online] Accessed: 2026-05-11}
}

@inproceedings{pandit2023combating,
  title={Combating robocalls with phone virtual assistant mediated interaction},
  author={Pandit, Sharbani and Sarker, Krishanu and Perdisci, Roberto and Ahamad, Mustaque and Yang, Diyi},
  booktitle={32nd USENIX Security Symposium (USENIX Security 23)},
  pages={463--479},
  year={2023}
}

@misc{cwa-federal-call-center-factsheet,
  author       = {{Communications Workers of America}},
  title        = {{Federal Call Center Factsheet}},
  howpublished = {\url{https://cwa-union.org/sites/default/files/federal_call_center_factsheet_.pdf}},
  note = {[Online] Accessed: 2026-05-11}
}

\end{document}